\documentclass[10pt,aps,prb,twocolumn]{revtex4}
\usepackage{graphicx}
\usepackage{epstopdf}
\usepackage{latexsym}
\usepackage{amssymb}
\usepackage{amsmath}
\usepackage{amsfonts}
\usepackage{subfigure}
\usepackage{bm}
\usepackage{multirow}

\newlength{\figwidth}

\begin{document}

\title{Magnetic and superconducting instabilities in a hybrid model of
itinerant/localized electrons for iron pnictides}
\author{Yi-Zhuang You$^1$, Fan Yang$^2$, Su-Peng Kou$^3$, Zheng-Yu Weng$^1$}
\affiliation{$^1$ Institute for Advanced Study, Tsinghua University - Beijing, 100084,
China\\
$^2$ Department of Physics, Beijing Institute of Technology - Beijing,
100081, China\\
$^3$ Department of Physics, Beijing Normal University - Beijing, 100875,
China}
\date{\today}

\begin{abstract}
We study a unified mechanism for spin-density-wave (SDW) and
superconductivity in a minimal model, in which itinerant electrons and local
moments coexist as previously proposed for the iron pnictides [EPL, 88,
17010 (2009)]. The phase diagram obtained at the mean field level is in
qualitative agreement with the experiment, which shows how the magnetic and
superconducting (SC) instabilities are driven by the critial coupling
between the itinerant/localized electrons. The spin and charge response
functions at the random phase approximation (RPA) level further characterize
the dynamical evolution of the system. In particular, the dynamic spin
susceptibility displays a Goldstone mode in the SDW phase, which evolves
into a gapped resonance-like mode in the superconducting phase. The latter
persists all the way into the normal state above $T_{c}$, where a strong
scattering between the itinerant electrons and local moments is restored, as
an essential feature of the model.
\end{abstract}

\maketitle

\section{Introduction}

High-temperature superconductivity found\cite{discovery1,discovery2} in the
iron pnictides has attracted an intensive attention in recent years.\cite%
{greene} The SC pairing of the electrons in these materials is less likely
to be mediated by phonons, as suggested by the local-density-approximation
(LDA) calculations\cite{phonon} as well as a variety of the experiments.\cite%
{greene} The proximity of the SC state to the SDW phase\cite{neutron} in the
phase diagram implies that the interplay between the magnetism and
superconductivity might play an important role in understanding the pairing
mechanism and other physical properties of the iron-based superconductors.
Combined with the high SC transition temperature, one finds an intriguing
resemblance between this family of materials and the cuprates, in which
superconductivity is generally believed of electronic origin.

Nevertheless the electrons in the iron pnictides are more itinerant than
those in the cuprates, especially in the magnetic phase, where the electrons
in the latter are localized due to a Mott transition to form local moments
which become antiferromagnetically (AF) ordered at low temperature. By
contrast, in the former, many different experiments have clearly
demonstrated the itinerancy of the electrons in the SDW phase, including the
multiple Fermi pockets as revealed by the angle-resolved photoemission
spectroscopy (ARPES) measurements,\cite{ARPES1,ARPES2,ARPES3,ARPES4,ARPES5}
which are consistent with the LDA calculations,\cite%
{SPM1,SPM2,LDA1,LDA4,LDA5} the quantum-oscillation,\cite{quntumnossillation}
the transport\cite{discovery1,discovery2,transport} as well as the optical
measurement.\cite{opt} Based on these experiments, one may reasonably view
the magnetic order in this system as an SDW order formed by the itinerant
electrons via Fermi-surface nesting.\cite{nesting1,LDA1,LDA5} From this
point of view, it is natural to conjecture that the superconducting pairing
is mediated by the collective magnetic fluctuations of the itinerant
electrons. A lot of theoretical efforts have been done along this line,
including the weak-coupling RPA theory,\cite{SPM1,SPM2,RPA1,RPA2} the
fluctuation-exchange approximation (FLEX),\cite{FLEX1,FLEX2} renormalization
group (RG),\cite{RG}, functional renormalization group (FRG)\cite%
{SPM3,FRG1,FRG2} and strong-coupling variational Monte-Carlo (VMC)\cite{VMC}
approaches. The sign change in the gap function between the electron and
hole pockets is predicted\cite{SPM1,SPM2,SPM3} as due to such unconventional
magnetic origin of superconductivity.

However, the magnetism in the iron pnictides has been also looked upon from
the strong coupling side, where the electrons are localized via a multiband
Mott transition forming local moments.\cite{yildirim,Si,yao,cenke,weng}
Indeed, the magnetic ordering phase observed in experiment can be also
described by utilizing a Heisenberg type $J_{1}$-$J_{2}$ model.\cite%
{neutronloc,loc} A study of the SC state based on a doped Mott-insulator
described by the $t$-$J_{1}$-$J_{2}$ model also yields\cite{tJ1J2} a
consistent pairing symmetry as compared to the experiment. This point of
view is further supported by the first-principle LDA calculations which
generally show the tendency for a large magnetic moment formation at low
doping.\cite{LDAmoment1,LDAmoment2,LDAmoment3,LDAmoment4,LDAmoment5} A
strong experimental support for the local moment picture comes from the
observation of a linear temperature dependent magnetic susceptibility in a
broad temperature regime above the N\'{e}el temperature.\cite%
{magsus1,magsus2,magsus3} A recent neutron-scattering measurement\cite%
{tranquada} in \textrm{FeTe}$_{0.35}\mathrm{Se}_{0.65}$ further observed a
substantial magnetic moment persisting up to $300$\textrm{K }above $T_{c}$
in the SC regime.

To reconcile the two aspects of the itinerant and localized electrons
exhibited in the iron-based superconductors, a minimal model was proposed in
Ref. \cite{hybrid model}, in which the itinerant electrons and the local
moments are conjectured to coexist, based on the multiband nature of the
system. Here the two separated degrees of freedom may be attributed\cite{wu}
to different 3d orbitals of the iron atoms, with the local moments formed
via an orbital-selective Mott transition.\cite{vojta} Recent dynamic
mean-field theory (DMFT) calculations\cite{DMFT1,DMFT2,DMFT3} lend numerical
support for the possible orbital-selective Mott transition in the iron-based
superconductors. Similar local-itinerant hybrid\ models with incorporating
the detailed orbital characters\cite{lv,5band} have been used to account for
the magnetic excitations observed in the iron pnictides.

The most essential feature in the model of coexistent itinerant/localized
electrons lies in the \emph{momentum match} between the two degrees of
freedom.\cite{hybrid model} Namely, the characteristic wave vector of the
magnetic correlation of the local moments, $\mathbf{Q}_{s}=(\pi ,0)$ or $%
(0,\pi )$, is commensurate with the typical momentum transfer between the
hole and electron pockets of the itinerant electrons. Consequently the
scattering between the two degrees of freedom, which are coupled by the
Hund's rule interaction, can get much enhanced in the normal state,
rendering the system intrinsically unstable towards either the magnetic or
superconducting ordering at low temperatures. At low doping, the
magnetically ordered state can be obtained\cite{hybrid model} as composed of
an SDW order of the itinerant electrons simultaneously locking with the
collinear magnetic order of the local moments at the same wave vector $%
\mathbf{Q}_{s}$. Here the nesting effect in the itinerant degrees of freedom
alone or a pure $J_{1}$-$J_{2}$ superexchange interaction in the local
moment part can be much weaker themselves in driving the magnetic
transition. For example, a perfect nesting of the Fermi surfaces can be
easily removed by adjusting the chemical potential\cite{hybrid model} or by
introducing a more realist band structure.\cite{5band} The pure collinear
antiferromagnetic (AF) ordering for the local moments can be also switched
off by setting\cite{hybrid model} $J_{1}=0$ and/or making the subsystem in
the disordered regime. But the hybrid system can nevertheless experience an
SDW phase transition due to the magnetic instability driven by the
above-mentioned critical coupling between the two subsystems. The residual
scattering between the itinerant and localized electrons will become much
reduced in the SDW state, which exhibits\cite{hybrid model,5band} a series
of magnetic and charge properties qualitatively consistent with the
iron-based superconductors at low doping.

However, the SDW instability is only one of the possible infrared
fixed-points of this minimal model at low doping. With the increase of
doping, the itinerant electrons can also form the Cooper pairs to gain the
interacting energy between the two subsystems. By doing so, the
superconducting energy gaps open up at the Fermi surfaces of the hole and
electron pockets, which also stabilize the system by reducing the residual
scattering between the itinerant and localized electrons. Generally the
origin of the SDW and SC states can be explored on equal footing in the
hybrid model of itinerant/localized electrons.

In this paper, a systematic evolution of the SDW and SC states with doping
is studied based on a simplified two-component model similar to the one
proposed in Ref. \cite{hybrid model}. Both the magnetic and SC instabilities
are found in this model at low temperatures, and a phase diagram is
determined at the mean-field level. In the SDW state, the dynamic spin
susceptibility at the RPA level shows that the magnetic excitations are
split into two branches composed of a low-lying Goldstone mode and a gapped
high-energy mode dominantly contributed by the local moments. The latter
generally gets severely broadened due to the strong scattering between the
itinerant electrons and local moments. On the other hand, the coherent
Goldstone mode in the magnetic phase will be replaced by a low-lying gapped
\textquotedblleft resonance-like\textquotedblright \ mode when the system
enters into the SC state, which can persist all the way to the
high-temperature normal state, consistent with the neutron-scattering
observations.\cite{spin resonance} The band renormalization is also studied
within the same framework. The results in this model study illustrate a
phenomenology consistent with the iron-based superconductors, in which
neither the Fermi surface nesting for the itinerant electrons nor the
superexchange interaction for the local moments play the direct role alone.
Rather these effects get strongly enhanced via the Hund's rule coupling
between the two subsystems with the momentum match. The latter effect makes
the system generically unstable against the SDW/SC orderings at different
doping and therefore provides a unified mechanism to understand both orders
appearing in the iron pnictides.

The remaining part of the paper is organized as follows. In section II, we
introduce the basic model, which includes a two-pocket description of the
itinerant electrons and a nonlinear $\sigma $-model description of the local
moments, which are coupled together by the Hund's rule coupling. In section
III, we present a mean-field calculation, which gives rise to a global phase
diagram with both the SDW and SC orders identified at different dopings.
Section IV is on the dynamic fluctuation beyond the mean-field
approximation, including the RPA calculations of the dynamic spin
susceptibility, the uniform magnetic susceptibility in different phases and
the band renormalization effect. Finally, the conclusion and discussion are
presented in section V.


\section{Model Description}

Our starting point is an itinerant-electron and local-moment hybrid model,
previously proposed\cite{hybrid model} to describe the iron-based
superconductors. We shall explore the emergent magnetism and
superconductivity in this highly simplified model, which is to be specified
below.

\subsection{Model Action}

The effective action includes an itinerant electron sector $S_{\text{it}}$,
a local moment sector $S_{\text{loc}}$, and a Hund's rule coupling term $S_{%
\text{H}}$ as follows
\begin{equation}
S_{\mathrm{eff}}=S_{\text{it}}+S_{\text{loc}}+S_{\text{H}}.  \label{Seff}
\end{equation}

\subsubsection{Itinerant electron}

We consider a simple two-pocket model for the itinerant electrons, whose
action reads
\begin{equation}
S_{\text{it}}=\sum_{k}c_{k}^{\dagger }(-i\omega +\xi _{\bm{k}})c_{k},
\end{equation}%
where $k=(i\omega ,{\bm{k}})$ is the (fermionic) momentum-frequency vector, and $%
c^{\dagger }=(c_{\Gamma \uparrow }^{\dagger },c_{\Gamma \downarrow
}^{\dagger },c_{M\uparrow }^{\dagger },c_{M\downarrow }^{\dagger })$
contains the creation operators of the itinerant electron for both spins ($%
\uparrow $ and $\downarrow $) and in the hole and electron pockets denoted
by $\Gamma $ and $M,$ respectively (see below). The momentum-frequency
summation stands for $\sum_{k}=\beta ^{-1}\sum_{i\omega }\sum_{\bm{k}}$,
where $\beta ^{-1}=k_BT$, and this convention will be adopted throughout this
work.

$\xi _{\bm{k}}$ is a $4\times 4$ matrix which determines the band
dispersion. It may be written as
\begin{equation}
\xi _{\bm{k}}=-\mu \rho _{0}-\epsilon _{\bm{k}}\rho _{3},  \label{eq:xi}
\end{equation}%
where $\rho _{i}\equiv \sigma _{i}\otimes \sigma _{0}$ is the $%
4\times 4$ matrix defined by the Kronecker product of the Pauli matrices $%
\sigma _{i}$ (where $\sigma _{0}$ is the identity $2\times 2$ matrix). $\mu $
is the chemical potential, and $\epsilon _{\bm{k}}={\bm{k}}%
^{2}/(2m)-\epsilon _{0}$ models a parabolic band dispersion with the
effective mass $m$ and energy shift $\epsilon _{0}$.

\begin{figure}[hb]
\centering
\subfigure[]{\includegraphics[height=0.12\textheight]{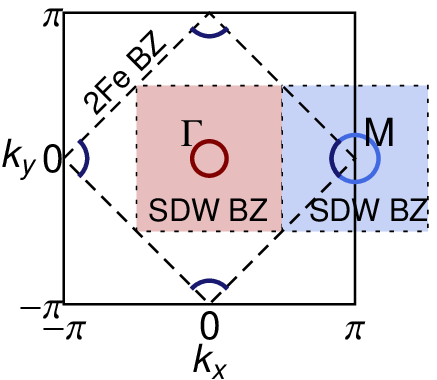}}
\quad \subfigure[]{\includegraphics[height=0.12%
\textheight]{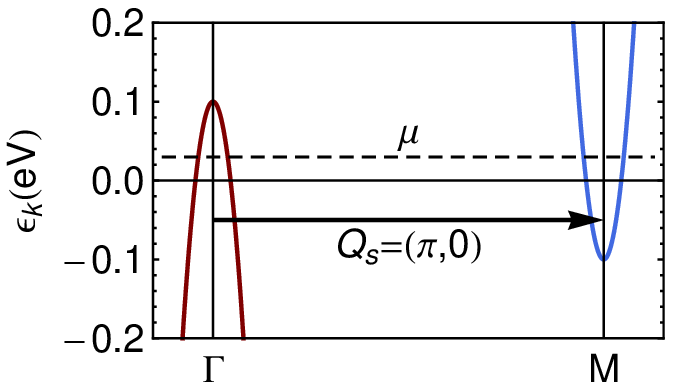}}
\caption{(Color online.) (a) The Fermi surface consists of a small hole
pocket around the $\Gamma$ point and a small electron pocket around $M$.
Both pockets locate around the origin in the SDW Brillouin zone (SDW BZ).
(b) Band structure. The nesting vector ${\bm{Q}_s}=(\protect \pi,0)$ connects
the two bands.}
\label{fig:Band}
\end{figure}

According to ARPES,\cite{ARPES1,ARPES2,ARPES3,ARPES4} magneto-oscillation
experiments,\cite{quntumnossillation} and LDA calculations,\cite%
{SPM1,SPM2,LDA1,LDA4,LDA5} both Fermi pockets are small (with $k_{F}\sim \pi
/10$) and shallow ($E_{F}\sim 0.1\text{eV}$). So by setting the parameters
at $m=0.1\text{eV}^{-1}$ and $\epsilon _{0}=0.1\text{eV}$, we can produce a
reasonable band structure as shown in Fig.\thinspace \  \ref{fig:Band}.

There are several aspects that we wish to comment on this simple two-pocket
model. Firstly, at $\mu =0$, the hole and the electron Fermi pockets
perfectly match under a momentum translation ${\bm{Q}}_{s}=(\pi ,0)$. In
real materials the nesting is not perfect even in the undoped case,\cite%
{ARPES1,ARPES2,ARPES3,ARPES4,SPM1,SPM2,LDA1,LDA4,LDA5} which may represented
by a small but finite $|\mu |$ here. Secondly, the detailed orbital
characters are neglected with the pockets taken to be rotationally
symmetric. Such a simple model can not account for some anisotropic
phenomenon such as the nodal SDW gap as discussed in Ref. \cite{nodalsdw}.
Thirdly, for simplicity in the present two-pocket band structure we focus on
the electron pocket around $(\pi ,0)$ point and the hole pocket around $(0,0)
$, which are connected by ${\bm{Q}}_{s}=(\pi ,0)$. The scattering between $%
(0,0)$ and $(0,\pi )$ pockets as connected by ${\bm{Q}}_{s}=(0,\pi )$ can be
similarly treated.

\subsubsection{Local moment}

\begin{figure}[hb]
\centering
\includegraphics[height=0.12\textheight]{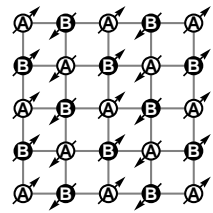}\newline
\caption{Each circle in the figure represents a Fe atom. The Fe lattice is
divided in to A and B sublattices. In each set of the lattice, the local moments
interact by the nearest AF coupling (next nearest in the original lattice).}
\label{fig:AB Sublattices}
\end{figure}

Consider an AF superexchange coupling $J_{2}$ bridged by the As ions between
the diagonal Fe sites. The local moments at the Fe sites may be divided into
two sets of sublattices, each is described by an AF Heisenberg model,
respectively. The effective Hamiltonian may be written as
\begin{equation}
H_{\text{loc}}=J_{2}\sum_{X=A,B}\sum_{\langle {\bm{r}}{\bm{r}}^{\prime
}\rangle \in X}M_{{\bm{r}}}\cdot M_{{\bm{r}}^{\prime }},
\end{equation}%
where $X=A,B$ labels the sites on different sublattices as shown in
Fig.\thinspace \ref{fig:AB Sublattices}, and $\langle {\bm{r}}{\bm{r}}%
^{\prime }\rangle $ denotes the nearest neighboring sites in the same
sublattice (or equivalently the next nearest neighboring sites in the
original lattice).

The local moment $M_{\bm{r}}$ will intrinsically fluctuate around the
characteristic momentum ${\bm{Q}}_{s}=(\pi ,0)$ or $(0,\pi )$. So we may
take $M_{\bm{r}}=Mn_{\bm{r}}e^{i{\bm{Q}}_{s}\cdot {\bm{r}}}$ with $n_{\bm{r}%
} $ a unit three-component real vector field, such that $n_{\bm{r}}$
fluctuates smoothly in the space, which will be convenient for further field
theoretical treatments.

The low-energy AF fluctuation of the local moment in each sublattice may be
described by a nonlinear $\sigma $-model. The corresponding action reads $S_{%
\text{loc}}=\sum_{X=A,B}S_{X}$, with
\begin{equation}
S_{X}=\frac{1}{4g}\sum_{q}n_{X,-q}\left( \nu ^{2}+c^{2}{\bm{q}}^{2}+\eta
^{2}\right) n_{X,q}-\frac{\eta ^{2}}{4g},
\end{equation}%
where $q=(i\nu ,{\bm{q}})$ denotes the (bosonic) momentum-frequency vector, and $%
n_{X,q}=\int_{0}^{\beta }d\tau \sum_{{\bm{r}}\in X}e^{i({\bm{q}}\cdot {\bm{r}%
}-\nu \tau )}n_{\bm{r}}(\tau )$ is Fourier-transformed from the field $n_{%
\bm{r}}$ in the $X$ sublattice. $\eta ^{2}$ is the Lagrangian multiplier
that enforces the unit condition $n_{\bm{r}}^{2}=1$. According to the above
definition, a long-wavelength limit at ${\bm{q}\rightarrow 0}$ actually
corresponds to the real momentum $\rightarrow $ ${\bm{Q}}_{s}.$

The coupling constant $g$ and spin wave velocity $c$ are related to the
Heisenberg $J_{2}$-model by $g=8J_{2}$, $c=4J_{2}M$ (with the Fe-Fe distance
taken as the unit).\cite{hybrid model} From the neutron-scattering
experiments,\cite{neutronloc} the typical spin wave velocity is around $0.3%
\text{eV}$, so that we set $c=0.3\text{eV}$, which in turn determines $%
J_{2}=0.093\text{eV}$ and $g=0.75\text{eV}$, assuming the magnetic moment $%
M=0.8$ per Fe atom.

By introducing a parallel field $n=(n_{A}+n_{B})/2$ and an antiparallel
field $\tilde{n}=(n_{A}-n_{B})/2$, the action can be further written as
\begin{equation}
\begin{split}
S_{\text{loc}}=& \frac{1}{2g}\sum_{q}n_{-q}(\nu ^{2}+c^{2}{\bm{q}}^{2}+\eta
^{2})n_{q} \\
+& \frac{1}{2g}\sum_{q}\tilde{n}_{-q}(\nu ^{2}+c^{2}{\bm{q}}^{2}+\eta ^{2})%
\tilde{n}_{q}-\frac{\eta ^{2}}{2g}.
\end{split}
\label{eq:S_loc}
\end{equation}

\subsubsection{Coupling term}

The itinerant electron spin $S_{\bm{r}}=Sc_{\bm{r}}^{\dagger }\sigma c_{%
\bm{r}}$ at site ${\bm{r}}$ can be coupled to the local moment $M_{\bm{r}}$
at the same site by a Hund's rule interaction
\begin{equation}
H_{\text{cp}}=-J_{H}\sum_{\bm{r}}M_{\bm{r}}\cdot S_{\bm{r}},
\end{equation}%
where $J_{H}$ denotes the strength of the effective Hund's rule coupling.
Fourier-transforming to the momentum-frequency space yields the following
action
\begin{equation}
S_{\text{cp}}=-J_{0}\sum_{k,q}n_{q}\cdot c_{k+q}^{\dagger }sc_{k},
\label{eq:S_cp}
\end{equation}%
where $J_{0}\equiv 2J_{H}MS$, $M$ is the magnitude of the local moment, and $%
S=1/2$ for the itinerant electron. In Eq. (\ref{eq:S_cp}), $s\equiv
(s_{1},s_{2},s_{3})$ with each $s_{i}$ ($i=1,2,3$) as a $4\times 4$ matrix
defined by $s_{i}\equiv \sigma _{1}\otimes \sigma _{i}$.

The itinerant electrons will couple to the parallel field $n$ which mainly
causes the scattering between the $\Gamma $ pocket and the M pocket at $(\pi
,0)$, and to the antiparallel field $\tilde{n}$ which mainly causes the
scattering between the $\Gamma $ and $(0,\pi )$ pockets.

\subsection{Mass-gap equation}

To simplify the calculation, we replace the Lagrangian multiplier $\eta ^{2}$
by its saddle point value. Evaluating the saddle point equation $\delta S_{%
\text{loc}}/\delta (\eta ^{2})=0$ at the one-loop level yields
\begin{equation}
-\frac{1}{2g}\left( 1-n_{0}^{2}\right) +3\sum_{q}\frac{1}{\nu _{n}^{2}+c^{2}{%
\bm{q}}^{2}+\eta ^{2}}=0,
\end{equation}%
where $n_{0}\equiv \langle n_{q=0}\rangle $ denotes the mean-field value of
the local moment.

Carrying out the momentum and frequency summations, $\eta $ can be solved
from the above equation as follows
\begin{equation}
\eta =\eta _{0}+\frac{2}{\beta }\ln \frac{1}{2}\left( 1+\sqrt{1+4e^{-\beta
\eta _{0}}}\right) ,  \label{eq:eta}
\end{equation}%
where $\eta _{0}$ is the solution of $\eta $ at zero temperature, which
reads
\begin{equation}
\eta _{0}=\frac{2\pi c^{2}}{3g_{c}}\frac{1-\gamma ^{2}}{2\gamma },
\end{equation}%
where $\gamma =(g_{c}/g)(1-n_{0}^{2})$. The parameter $g_{c}=2\pi
c/(3\Lambda )$ is introduced to replace the momentum cutoff $\Lambda $ which
is needed to control the convergence of the momentum summation.

The physical meaning of $\eta $ is the mass gap of the bare spin wave
excitation in the local moment sector. If $g>g_{c}$, the mass gap will
remain finite at zero temperature, indicating a disordered state with only
short-range magnetic ordering for the pure local moment degrees of freedom.
We shall see that due to the strong coupling between the itinerant electrons
and local moments, a true SDW order can be still induced, even if the bare
local moments are in a disordered regime and do not order at $T=0$ by
themselves$.$

\subsection{Propagators}

The itinerant electron single-particle propagator is denoted by $G(k)\equiv
-\langle c_{k}c_{k}^{\dagger }\rangle $, and represented by the arrowed line %
\settowidth{\figwidth}{\includegraphics[scale=0.75]{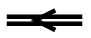}} %
\parbox{\figwidth}{\includegraphics[scale=0.75]{G.eps}} in the Feynman
diagram. $G(k)$ is a $4\times 4$ matrix as $c_{k}^{\dagger }$ has 4
components (2 pockets $\times $ 2 spins). The propagator for the local
moment is denoted by $D(q)\equiv -\langle n_{-q}n_{q}\rangle $, and
represented by the dashed line \settowidth{\figwidth}{%
\includegraphics[scale=0.75]{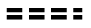}} \parbox{\figwidth}{%
\includegraphics[scale=0.75]{D.eps}} in the Feynman diagram. $D(q)$ is a $%
3\times 3$ matrix as $n_{q}$ has 3 components.

The double lines in the Feynman diagram represent the dressed propagators,
while the single lines, such as \settowidth{\figwidth}{%
\includegraphics[scale=0.75]{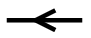}} \parbox{\figwidth}{%
\includegraphics[scale=0.75]{G0.eps}} and \settowidth{\figwidth}{%
\includegraphics[scale=0.75]{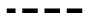}} \parbox{\figwidth}{%
\includegraphics[scale=0.75]{D0.eps}}, denote the bare propagators, $G_{0}$
and $D_{0},$ respectively. The latter can be read out directly from the
actions, $S_{\text{it}}$ and $S_{\text{loc}}$, as
\begin{equation}
G_{0}(k)=(i\omega -\xi _{{\bm{k}}})^{-1}=\frac{(i\omega +\mu )\rho
_{0}-\epsilon _{{\bm{k}}}\rho _{3}}{(i\omega +\mu )^{2}-\epsilon _{{\bm{k}}%
}^{2}},  \label{eq:G_0}
\end{equation}%
and%
\begin{equation}
D_{0,ii^{\prime }}(q)=\frac{-g\delta _{ii^{\prime }}}{\nu ^{2}+c^{2}{\bm{q}}%
^{2}+\eta ^{2}}.  \label{eq:D_0}
\end{equation}

\section{Mean-Field Phase Diagram}

In the effective action Eq. (\ref{Seff}), even though the two subsystems,
i.e., the local moments and itinerant electrons, may not be in a magnetic or
SC ground state separately, the whole system coupled together will
experience intrinsic magnetic and SC instabilities, which are studied on
equal footing below at the mean-field level.

\subsection{SDW phase}

The SDW mean-field equation can be deduced from the following Dyson
equation for the self-consistent Hartree approximation
\begin{equation}
\settowidth{\figwidth}{\includegraphics[scale=0.75]{G.eps}}%
\parbox{\figwidth}{\includegraphics[scale=0.75]{G.eps}}=\settowidth{%
\figwidth}{\includegraphics[scale=0.75]{G0.eps}}\parbox{\figwidth}{%
\includegraphics[scale=0.75]{G0.eps}}+\settowidth{\figwidth}{%
\includegraphics[scale=0.75]{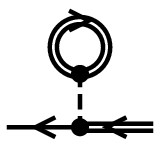}}\parbox{\figwidth}{%
\includegraphics[scale=0.75]{G0SdG.eps}}\text{ \  \ },
\end{equation}%
or
\begin{equation}
G(k)=G_{0}(k)+G_{0}(k)\Sigma _{d}G(k),  \label{eq:SDW dynson}
\end{equation}%
where the Hartree self-energy is given by
\begin{equation}
\Sigma _{d}=J_{0}^{2}D_{0,ii^{\prime }}(q=0)s_{i}\sum_{k^{\prime }}\text{Tr}%
\left[ G\left( k^{\prime }\right) s_{i^{\prime }}\right] .
\label{eq:Sigma d}
\end{equation}

On the other hand, the Hartree energy is related to the local moment mean
field by
\begin{equation}
\Sigma _{d}=-J_{0}n_{0i}s_{i},  \label{eq:Sigma d form}
\end{equation}%
where $n_{0i}$ is the $i$th component of $\langle n_{q=0}\rangle $. This can
be seen by separating out the $q=0$ term $-J_{0}\sum_{k}n_{0}\cdot
c_{k}^{\dagger }sc_{k}$ from the summation in the coupling term
Eq.\thinspace \eqref{eq:S_cp}. This term indicates that the local moment
mean field $n_{0}$ affects the itinerant electron Hamiltonian by adding the
self-energy $-J_{0}n_{0}\cdot s$, which should just be identified as the
Hartree energy. The consistency between Eq.\thinspace \eqref{eq:Sigma d} and
Eq.\thinspace \eqref{eq:Sigma d form} leads to the SDW mean field equation.
To show this, we first take Eq.\thinspace \eqref{eq:Sigma d form} and
evaluate the propagator from Eq.\thinspace \eqref{eq:SDW dynson}
\begin{equation}
G(k)=\frac{(i\omega +\mu )\rho _{0}-\epsilon _{{\bm{k}}}\rho
_{3}-J_{0}n_{0i}s_{i}}{(i\omega +\mu )^{2}-\epsilon _{{\bm{k}}%
}^{2}-J_{0}^{2}n_{0}^{2}}.  \label{eq:G_SDW}
\end{equation}%
Substituting into Eq.\thinspace \eqref{eq:Sigma d} yields
\begin{equation}
\Sigma _{d}=\frac{4gJ_{0}^{3}n_{0i}s_{i}}{\eta ^{2}}\sum_{k}\frac{1}{%
(i\omega +\mu )^{2}-\epsilon _{{\bm{k}}}^{2}-J_{0}^{2}n_{0}^{2}}.
\end{equation}%
After Matsubara frequency summation,
\begin{equation}
\Sigma _{d}=-\frac{4gJ_{0}^{3}n_{0i}s_{i}}{\eta ^{2}}\sum_{{\bm{k}}}\frac{%
\sinh \beta E_{{\bm{k}}}}{2E_{{\bm{k}}}(\cosh \beta \mu +\cosh \beta E_{{%
\bm{k}}})},  \label{eq:Sigma d expression}
\end{equation}%
where $E_{\bm{k}}^{2}=\epsilon _{\bm{k}}^{2}+J_{0}^{2}n_{0}^{2}$. Comparing
with Eq.\thinspace \eqref{eq:Sigma d form}, we arrive at the SDW mean-field
equation
\begin{equation}
\frac{4gJ_{0}^{2}}{\eta ^{2}}\sum_{{\bm{k}}}\frac{\sinh \beta E_{{\bm{k}}}}{%
2E_{{\bm{k}}}(\cosh \beta \mu +\cosh \beta E_{{\bm{k}}})}=1,
\label{eq:SDW mean field}
\end{equation}%
from which the SDW order parameter $n_{0}$ can be determined for the given
chemical potential and temperature.


\subsection{SC Phase}

To deal with the SC order, we introduce the abnormal propagator for the
itinerant electrons: $F(k)\equiv -\langle c_{k}c_{-k}\rangle $, represented
by a line with two arrows heading in opposite directions \settowidth{%
\figwidth}{\includegraphics[scale=0.75]{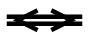}} \parbox{\figwidth}{%
\includegraphics[scale=0.75]{F.eps}}.

The SC mean-field equations are equivalent to the following Dyson equations
of self-consistent Fock approximation\cite{BCS},
\begin{equation}
\begin{split}
\settowidth{\figwidth}{\includegraphics[scale=0.75]{G.eps}}%
\parbox{\figwidth}{\includegraphics[scale=0.75]{G.eps}}& =%
\settowidth{\figwidth}{\includegraphics[scale=0.75]{G0.eps}}%
\parbox{\figwidth}{\includegraphics[scale=0.75]{G0.eps}}+%
\settowidth{\figwidth}{\includegraphics[scale=0.75]{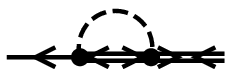}}%
\parbox{\figwidth}{\includegraphics[scale=0.75]{GG0SpF.eps}}, \\
\settowidth{\figwidth}{\includegraphics[scale=0.75]{F.eps}}%
\parbox{\figwidth}{\includegraphics[scale=0.75]{F.eps}}& =%
\settowidth{\figwidth}{\includegraphics[scale=0.75]{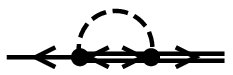}}%
\parbox{\figwidth}{\includegraphics[scale=0.75]{GG0SpGG.eps}}.
\end{split}%
\end{equation}%
The above diagrams correspond to
\begin{equation}
\begin{split}
G(k)& =G_{0}(k)-G_{0}(k)\Sigma _{p}(k)F(-k), \\
-F(k)& =G_{0}(k)\Sigma _{p}(k)G(-k),
\end{split}
\label{eq:SC dynson}
\end{equation}%
where the pairing energy is
\begin{equation}
\Sigma _{p}(k)=-J_{0}^{2}\sum_{k^{\prime }}D_{0,ii^{\prime }}\left(
k-k^{\prime }\right) s_{i}F\left( k^{\prime }\right) \left( -s_{i^{\prime
}}^{\intercal }\right) .  \label{eq:Sigma p}
\end{equation}

Mediated by the local moment fluctuation, the effective interaction between
the itinerant electrons is described by the following action
\begin{equation}
S_{\text{int}}=\sum_{k,k^{\prime },p}c_{k+p}^{\dagger }c_{-k}^{\dagger
}\Gamma (k-k^{\prime })c_{-k^{\prime }}c_{k^{\prime }+p},
\end{equation}%
where the vertex function reads
\begin{equation}
\Gamma (q)=-\settowidth{\figwidth}{\includegraphics[scale=0.75]{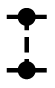}}%
\parbox{\figwidth}{\includegraphics[scale=0.75]{int.eps}}=\frac{J_{0}^{2}}{2}%
D_{0,ii^{\prime }}(q)s_{i}\otimes s_{i^{\prime }}.
\end{equation}%
The eigenvalue of the kernel function $\Gamma (k-k^{\prime })$ stands for
the effective pairing energy of the Cooper pair, whose form factor is
given by the corresponding eigenvector. The most negative eigenvalue (hence
the strongest pairing attraction) is found in the
spin-singlet intra-pocket pairing channel with $s^{\pm }$-wave symmetry. In
fact, simply by diagonalizing the matrix $s_{i}\otimes s_{i}$, it is easy to
show that the greatest eigenvalues belong to the spin-singlet parings with
opposite sign between the electron and the hole pocket. Among the spin-singlet pairing channels,
the inter-pocket pairing would lead to the pocket-singlet which requires the
gap function to be of $p$-wave symmetry and is not able to fully gap the
Fermi surface, and thus the intra-pocket $s^{\pm }$-wave pairing remains
most favorable.

By introducing a $4\times 4$ matrix $d^{\dagger }\equiv \sigma _{3}\otimes
(i\sigma _{2})$, the $s^{\pm }$-wave paring operator can be simply denoted
as
\begin{equation}
\begin{split}
c_{k}^{\dagger }d^{\dagger }c_{-k}^{\dagger }=& (c_{k\Gamma \uparrow
}^{\dagger }c_{-k\Gamma \downarrow }^{\dagger }-c_{k\Gamma \downarrow
}^{\dagger }c_{-k\Gamma \uparrow }^{\dagger }) \\
-& (c_{kM\uparrow }^{\dagger }c_{-kM\downarrow }^{\dagger }-c_{kM\downarrow
}^{\dagger }c_{-kM\uparrow }^{\dagger }).
\end{split}%
\end{equation}%
Therefore one may assume the pairing energy to take the same form
\begin{equation}
\Sigma _{p}(k)=\Delta (k)d,  \label{eq:Sigma p form}
\end{equation}%
with symmetric gap function $\Delta (k)$, i.e. $\Delta (-k)=\Delta (k)$.
Although the other pairing modes may also appear in $\Sigma _{p}$, they are
all omitted to simplify the derivation. In general, an $s^{++}$-wave pairing
can be also induced from $s^{\pm }$-wave pairing if the two Fermi pockets
are no longer symmetric in size (i.e. with a finite chemical potential $\mu $
in our model). However, according to our calculation, the $s^{\pm }$-wave
will be the dominant component persisting up to larger $\mu $.

Then from Eq.\thinspace \eqref{eq:SC dynson}, one can find the solution of $%
G(k)$ and $F(k)$,
\begin{equation}
G(k)=\frac{Z(k)\rho _{0}-E(k)\rho _{3}}{Z(k)^{2}-E(k)^{2}},  \label{eq:G_SC}
\end{equation}%
where
\begin{equation}
\begin{split}
Z(k)& =\frac{(i\omega -\mu )^{2}-\epsilon _{{\bm{k}}}^{2}-\Delta (k)^{2}}{%
(i\omega -\mu )^{2}-\epsilon _{{\bm{k}}}^{2}}(i\omega -\mu )+2\mu , \\
E(k)& =\frac{(i\omega -\mu )^{2}-\epsilon _{{\bm{k}}}^{2}-\Delta (k)^{2}}{%
(i\omega -\mu )^{2}-\epsilon _{{\bm{k}}}^{2}}\epsilon _{{\bm{k}}};
\end{split}%
\end{equation}%
and
\begin{equation}
F(k)=f(k)\Delta (k)d,  \label{eq:F}
\end{equation}%
where
\begin{equation}
f(k)=\frac{1}{2}\sum_{\varsigma =\pm 1}\frac{1}{(i\omega )^{2}-(\epsilon _{{%
\bm{k}}}+\varsigma \mu )^{2}-\Delta (k)^{2}}.
\end{equation}%
Here we have projected out the components other than $s^{\pm }$-wave in the
solution of $F(k)$ as noted above. Substituting the above solutions into
Eq.\thinspace \eqref{eq:Sigma p} yields
\begin{equation}
\Sigma _{p}(k)=3J_{0}^{2}\sum_{k^{\prime }}D_{0,11}\left( k-k^{\prime
}\right) f\left( k^{\prime }\right) \Delta \left( k^{\prime }\right) d.
\end{equation}%
Comparing with Eq.\thinspace \eqref{eq:Sigma p form}, we arrive at the SC
mean-field equation
\begin{equation}
\Delta (k)=3J_{0}^{2}\sum_{k^{\prime }}D_{0,11}\left( k-k^{\prime }\right)
f\left( k^{\prime }\right) \Delta \left( k^{\prime }\right) .
\end{equation}%
This equation can be solved by numerical approach.

To proceed with analytic analysis, we omit the $k$-dependence of $\Delta (k)$
and replace it by a constant $\Delta $. We also approximate $%
D_{0}(k-k^{\prime })$ by its average value $\langle D_{0}\rangle $ at zero
frequency around the Fermi surface,
\begin{equation}
\left \langle D_{0}\right \rangle =-g\left \langle \frac{1}{c^{2}({\bm{k}}-{%
\bm{k}}^{\prime })^{2}+\eta ^{2}}\right \rangle _{k,k^{\prime }\in \text{FS}%
}=-\frac{g}{\eta _{\text{eff}}^{2}}.
\end{equation}%
One finds $\eta _{\text{eff}}^{2}=\eta \left( \eta
^{2}+4c^{2}k_{F}^{2}\right) ^{1/2}$ with $k_{F}^{2}=2m\epsilon _{0}$. Then
the SC mean-field equation becomes
\begin{equation}
1=-\frac{3gJ_{0}^{2}}{\eta _{\text{eff}}^{2}}\sum_{k}f(k),
\end{equation}%
which, after the Matsubara frequency summation, yields
\begin{equation}
1=\frac{3gJ_{0}^{2}}{2\eta _{\text{eff}}^{2}}\sum_{\varsigma =\pm 1}\sum_{{%
\bm{k}}}\frac{1}{2E_{\varsigma ,{\bm{k}}}}\tanh \frac{\beta E_{\varsigma ,{%
\bm{k}}}}{2},  \label{eq:SC mean field}
\end{equation}%
where $E_{\varsigma ,{\bm{k}}}^{2}=\left( \epsilon _{{\bm{k}}}+\varsigma \mu
\right) ^{2}+\Delta ^{2}$, which determines the SC gap $\Delta $.


\subsection{Phase Diagram}

\begin{figure}[tbp]
\includegraphics[height=0.17\textheight]{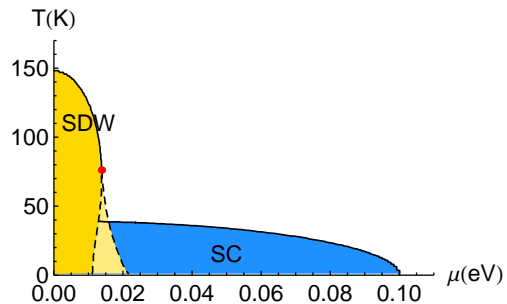}\newline
\caption{(Color online.) The phase diagram calculated according to the mean
field equations. The solid lines denote the second order phase transition
boundaries, and the dashed lines denote the first order phase transition
boundaries. The red point is a critical point. The SDW and SC phase
boundaries are calculated independently, and the overlap region does not necessarily
imply the coexistence of two orders.}
\label{fig:Phase Diagram}
\end{figure}

By solving self-consistently the two mean-field equations, Eqs.\thinspace %
\eqref{eq:SDW mean field} and \thinspace \eqref{eq:SC mean field}, the phase
diagram of the SDW and SC phases can be determined as shown in
Fig.\thinspace \ref{fig:Phase Diagram}. In the following, we specify the
choice of the parameters in the model.

It is noted that there are three particular points in the phase diagram that
can be calculated analytically based on the mean-field equations. They are
the SDW critical temperature $T_{\text{SDW}}^{0}$ at $\mu =0$, the SC
critical temperature $T_{c}^{0}$ at $\mu =0$ (if not consider SDW), and the
chemical potential $\mu _{\text{SDW1}}$ at which the SDW order disappears at
zero temperature. The formulae read
\begin{equation}
T_{\text{SDW}}^{0}=1.13T_{0}e^{-1/(N_{F}^{0}V_{\text{SDW}})},
\label{eq:TSDW}
\end{equation}%
\begin{equation}
T_{c}^{0}=1.13T_{0}e^{-1/(N_{F}^{0}V_{\text{SC}})},  \label{eq:TSC}
\end{equation}%
and%
\begin{equation}
\mu _{\text{SDW1}}=T_{0}e^{-1/(N_{F}^{0}V_{\text{SDW}})},  \label{eq:muSDW1}
\end{equation}%
where $N_{F}^{0}=m/(2\pi )$ is the density of state at the Fermi energy $\mu
=0$, $T_{0}\equiv \epsilon _{0}(\Lambda ^{2}/(2m\epsilon _{0})-1)^{1/2},$ $%
V_{\text{SDW}}=4gJ_{0}^{2}/\eta ^{2}$ and $V_{\text{SC}}=3gJ_{0}^{2}/\eta _{%
\text{eff}}^{2}.$ To be overall comparable to the experiments, we take $T_{%
\text{SDW}}^{0}\simeq 150$K and $T_{c}^{0}\simeq 40$K and make use of
Eqs.\thinspace \eqref{eq:TSDW} and \thinspace \eqref{eq:TSC} to fix the
model parameters: $g_{c}=0.53$eV and $J_{0}=0.39$eV. Consequently, with all
the basic parameters in our model given, the phase diagram is determined
numerically as a function of the chemical potential in Fig.\thinspace \ref%
{fig:Phase Diagram}.

\begin{figure}[tbp]
\includegraphics[height=0.13\textheight]{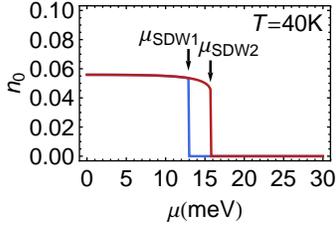}\newline
\caption{The SDW order as a function of chemical potential at $T=40K$,
showing the hysteresis loop of the first ordered transition with two
distince transition chemical potential.}
\label{fig:2 SDW Transitions}
\end{figure}

Figure \thinspace \ref{fig:Phase Diagram} shows a critical point at the SDW
phase boundary, located at $\mu ^{\ast }=0.014\text{eV}$ and $T^{\ast }=76%
\text{K}$, where the second-order phase transition boundary splits into two
first-order phase transition boundary lines. Hence at $T=0,$ there exist two
critical chemical potentials for the SDW transition: $\mu _{\text{SDW1}%
}=0.011\text{eV}$ and $\mu _{\text{SDW2}}=0.022\text{eV,}$ respectively, as
shown in Fig.\thinspace \  \ref{fig:2 SDW Transitions}. A similar first-order
transition phenomenon has been also reported\cite{critical point} in some
other theoretical approach based on a pure itinerant model.

Figure\thinspace \ref{fig:Phase Diagram} shows that the SC critical
temperature is not sensitive to the chemical potential, in contrast to the
SDW state, because the pairing mechanism of the present model is not
sensitive to the Fermi surface nesting condition. On the other hand, the $%
s^{\pm }$-wave superconductivity does require finite Fermi surface densities
of states in both hole and electron pockets to support the intra-pocket
pairing. That explains why $T_{c}$ eventually vanishes at $\mu =0.1\text{eV}$
when the Fermi level touches one of the band bottoms of the hole/electron
pockets. It is noted that beyond $\mu =0.1$eV, superconductivity of other
types of pairing symmetry is still possible in the present model, which will
involve the intra electron or hole pocket pairing and require some
incommensurate AF fluctuations of the local moment away from $\mathbf{Q}_{s}$%
.

Therefore, the high-temperature normal state can be regarded as an unstable
fixed point state in the present model, in which the itinerant electrons
scatter strongly with the local moments due to the momentum match at $%
\mathbf{Q}_{s}$ and the model defines the relevant degrees of freedom that
render the system flow into either an SDW or SC ordered phase, depending on
doping, as the temperature lowers. The mean-field equations Eqs.\thinspace %
\eqref{eq:SDW mean field} and \eqref{eq:SC mean field} describe,
respectively, how the SDW and SC orders emerge from such a normal state,
with the phase diagram in qualitative agreement with the iron
superconductors.\cite{greene}

Finally, we point out that although Fig. \ref{fig:Phase Diagram} suggests
that the SDW and SC phases may coexist at low doping, in mapping out the
phase diagram in the figure, only the maximal temperature of $T_{\text{SDW}}$
and $T_{c}$ is shown at a given $\mu $ with assuming the vanishing of the
other order in the mean-field equations. In other words, in order to
determine the coexistent SC state \emph{inside} the SDW regime, one needs to
further incorporate the detailed competition of the two orders into the
self-consistent mean-field equations, which can be straightforwardly done by
generalizing the above formulation. But this is not considered here not only
for the sake of simplicity, but also because we wish to emphasize that the
mutual interplay between the SDW and SC orders are \emph{not} essential in
driving their own formations in our model.


\section{Dynamic fluctuations}

The interaction between the itinerant and localized electrons has played a
crucial role in resulting in the SDW and SC states, as described by the
global diagram in Fig. \ref{fig:Phase Diagram}. In the following we further
investigate the evolution of dynamic fluctuations beyond the mean-field
approximation in these phases.

\subsection{Dynamic spin susceptibility}

To study the low-energy spin dynamics around the SDW wave vector $\mathbf{Q}%
_{s}$, we first consider the RPA correction to the propagator $D(q)$ of the
local moment by the Dyson equation
\begin{equation}
\settowidth{\figwidth}{\includegraphics[scale=0.75]{D.eps}}%
\parbox{\figwidth}{\includegraphics[scale=0.75]{D.eps}}=\settowidth{%
\figwidth}{\includegraphics[scale=0.75]{D0.eps}}\parbox{\figwidth}{%
\includegraphics[scale=0.75]{D0.eps}}+\settowidth{\figwidth}{%
\includegraphics[scale=0.75]{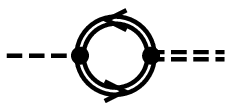}}\parbox{\figwidth}{%
\includegraphics[scale=0.75]{D0PD.eps}},
\end{equation}%
or
\begin{equation}
D(q)=D_{0}(q)+D_{0}(q)\Pi (q)D(q),  \label{eq:Dyson RPA}
\end{equation}%
which is solved formally as
\begin{equation}
D(q)=\frac{D_{0}(q)}{1-D_{0}(q)\Pi (q)},  \label{eq:D}
\end{equation}
where the RPA bubble $\Pi $ is given by
\begin{equation}
\Pi _{ii^{\prime }}(q)=-\settowidth{%
\figwidth}{\includegraphics[scale=0.75]{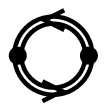}}\parbox{\figwidth}{%
\includegraphics[scale=0.75]{P.eps}}=J_{0}^{2}\sum_{k}\text{Tr}\left[ s_{i}G(k)s_{i^{%
\prime }}G(k+q)\right] .
\end{equation}%
Here as the propagator of the itinerant electron, the particular form of $%
G(k)$ depends on the mean field states: in the normal state, it takes the
form of Eq.\thinspace \eqref{eq:G_0}; in the SDW state, it is given by
Eq.\thinspace \eqref{eq:G_SDW}. While in the SC state, the contribution from
the abnormal propagator should be included as well
\begin{equation}
\begin{split}
\Pi _{ii^{\prime }}(q)=&-\settowidth{%
\figwidth}{\includegraphics[scale=0.75]{P.eps}}\parbox{\figwidth}{%
\includegraphics[scale=0.75]{P.eps}}-\settowidth{%
\figwidth}{\includegraphics[scale=0.75]{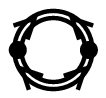}}\parbox{\figwidth}{%
\includegraphics[scale=0.75]{PF.eps}}\\
=& J_{0}^{2}\sum_{k}\text{Tr}\left[ s_{i}G(k)s_{i^{%
\prime }}G(k+q)\right] \\
+& J_{0}^{2}\sum_{k}\text{Tr}\left[ s_{i}F(k)(-s_{i^{\prime }}^{\intercal
})F^{\dagger }(k+q)\right] ,
\end{split}%
\end{equation}%
where Eqs.\thinspace \eqref{eq:G_SC} and \thinspace \eqref{eq:F} are used.

By noting $-J_{0}^{-2}\Pi (q)$ represents the spin susceptibility of the
itinerant electrons at the mean-field level, one can similarly write down
the spin susceptibility of the itinerant electrons at the RPA level, and
finally obtain the following total spin susceptibility
\begin{equation}
\chi (q)=-\frac{D_{0}(q)+J_{0}^{-2}\Pi (q)}{1-D_{0}(q)\Pi (q)}.
\end{equation}

\begin{figure}[hb]
\centering
\subfigure[]{\includegraphics[height=0.11\textheight]{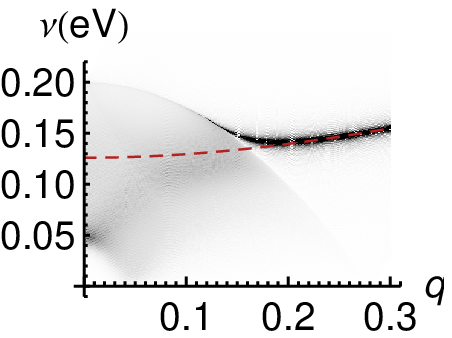}%
\label{fig:INS0}} \quad \subfigure[]{\includegraphics[height=0.11%
\textheight]{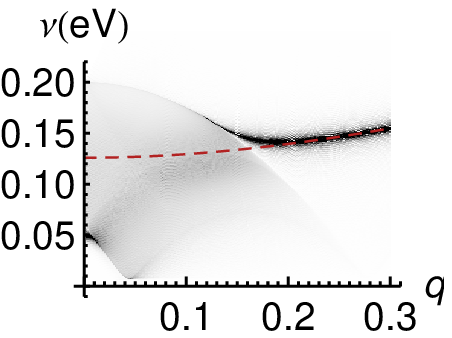}\label{fig:INSSC}}\newline
\subfigure[]{\includegraphics[height=0.11\textheight]{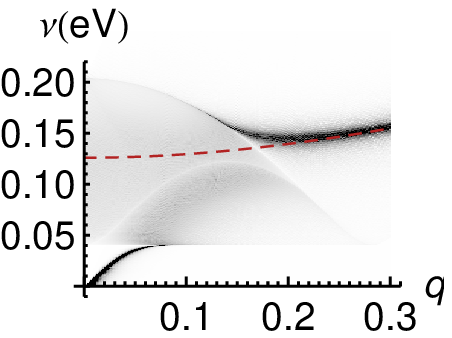}%
\label{fig:INSSDW1}} \quad \subfigure[]{\includegraphics[height=0.11%
\textheight]{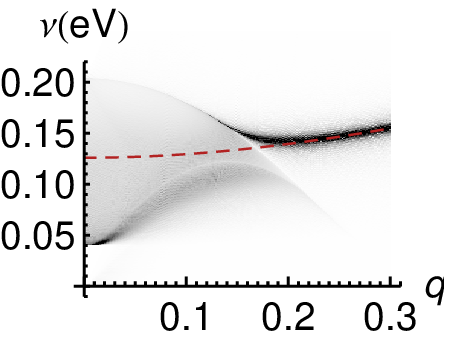}\label{fig:INSSDW3}}\newline
\caption{(Color online.) The calculated spectral function of dynamic spin susceptibility. Darker shade indicates
higher intensity. (a) in the normal phase ($\protect \mu=0.03\text{eV}$, $T=50%
\text{K}$). (b) in the SC phase ($\protect \mu=0.03\text{eV}$, $T=30\text{K}$%
). (c) and (d) are both in the SDW phase ($\protect \mu=0.00\text{eV}$, $T=80%
\text{K}$). (c) shows the spectrum of transverse fluctuation, and (d) is for the longitudinal fluctuation. The red dashed curve marks out the bare spin wave dispersion.}
\label{fig:INS}
\end{figure}

The inelastic neutron-scattering spectroscopy (INS) can measure the dynamic
spin susceptibility as the imaginary part of $\chi (q)$, $-\text{Im}\chi
(\nu +i0_{+},{\bm{q}})$, obtained after the Wick rotation $i\nu \rightarrow
\nu +i0_{+}$, which is presented in Fig.\thinspace \  \ref{fig:INS} in
different phases (see the figure caption for the details).

In the normal state, as Fig.\thinspace \ref{fig:INS0} shows, the spectrum of
the local moment fluctuation becomes very fuzzy when it immerses into the
continuum of the itinerant electrons as indicated by the dome-shaped shadow
area around $q\lesssim 0.2$. For comparison, the bare dispersion of the local moment spin wave
is marked out by the dashed curve (which is gapped
as in a disordered regime of the nonlinear $\sigma $-model as noted before).
The smearing of the spectrum is clearly the result of the strong scattering
between the itinerant and local moment in the region around $\mathbf{Q}_{s}$%
. At $q\simeq 0$, a hot spot at the frequency slightly below $2\mu $
can be seen in Fig.\thinspace \ref{fig:INS0}. This is because to
create a spin flip at $q=0$ involves a pair of electron and hole excitations at the $%
\Gamma $ and $M$ pockets respectively which costs at least energy $2\mu $ to go across
the Fermi surface, and on the other hand, the gap of the local moment
fluctuation is higher than this energy such that the scattering diminishes.
As a matter of fact, such a \textquotedblleft
resonance-like\textquotedblright \ mode becomes even sharper in the SC phase
(Fig.\thinspace \ref{fig:INSSC}) simply due to the further reduction of the
scattering with opening the SC gap. It may account for the \textquotedblleft
resonance mode\textquotedblright \ found in the INS experiment,\cite{spin
resonance} which indeed persists all the way to the normal state.

In the SDW state at low doping, inside the SDW gap of the itinerant
electrons, the fuzzy continuum is replaced by some emergent collective
modes. The transverse spin fluctuations (in the directions perpendicular to
the ordering direction) become the gapless Goldstone modes (Fig.\thinspace %
\ref{fig:INSSDW1}), which is consistent with the previous RPA calculation
using a more complicated five-band model for the itinerant electrons.\cite%
{5band} On the other hand, the longitudinal fluctuation (along the ordering direction) remains gapped as shown in Fig.\thinspace \ref{fig:INSSDW3}.

The existence of the Goldstone mode can be proven rigorously at the RPA
level. Since the RPA bubble in the SDW phase has a rather simple expression
at zero frequency and momentum
\begin{equation}
\begin{split}
\Pi _{ii^{\prime }}(0)=-4J_{0}^{2}\sum_{{\bm{k}}}& (\delta _{ii^{\prime }}-%
\frac{J_{0}^{2}n_{0i}n_{0i^{\prime }}}{E_{{\bm{k}}}^{2}}) \\
& \frac{\sinh \beta E_{{\bm{k}}}}{2E_{{\bm{k}}}(\cosh \beta \mu +\cosh \beta
E_{{\bm{k}}})}.
\end{split}%
\end{equation}%
Let us suppose that the SDW ordering is along the 3rd direction in the spin
space, i.e. $n_{0,1}=n_{0,2}=0$ and $n_{0,3}\neq 0$. Then $\Pi _{11}(0)=\Pi
_{22}(0)\leq \Pi _{33}(0)<0$, and by comparing with Eq.\thinspace %
\eqref{eq:Sigma d expression} and referring to Eq.\thinspace \eqref{eq:D_0},
it is recognized that
\begin{equation}
\Sigma _{d}=\frac{gJ_{0}n_{0i}s_{i}}{\eta ^{2}}\Pi
_{11}(0)=-J_{0}n_{0i}s_{i}D_{0,11}(0)\Pi _{11}(0).
\end{equation}%
As the self-energy $\Sigma _{d}$ is determined self-consistently from the
SDW mean-field equation $\Sigma _{d}=-J_{0}n_{0i}s_{i}$, at the mean-field
saddle point, we have $D_{0,11}(0)\Pi _{11}(0)=1$, which leads to a pole of $%
D_{11}$ at $q=0$ according to $D=(1-D_{0}\Pi )^{-1}D_{0}$, proving the
existence of a zero energy collective mode, i.e. the Goldstone mode. The
same argument applies for the $D_{22}$ component as well. Also taken into account the fact that $D_{12}=0$, it can be concluded that there are
two Goldstone modes, both are in the transverse directions. As for the $%
D_{33}$ component, since $\Pi _{33}\geq \Pi _{11}$ such that $D_{0,33}(0)\Pi
_{33}(0)\leq 1$ (note that $D_{0}(0)$ is negative), no pole can appear at $%
q=0$ in general, meaning that the longitudinal mode is still gapped.


\subsection{Uniform Susceptibility}

\begin{figure}[hb]
\centering
\subfigure[]{\includegraphics[height=0.11\textheight]{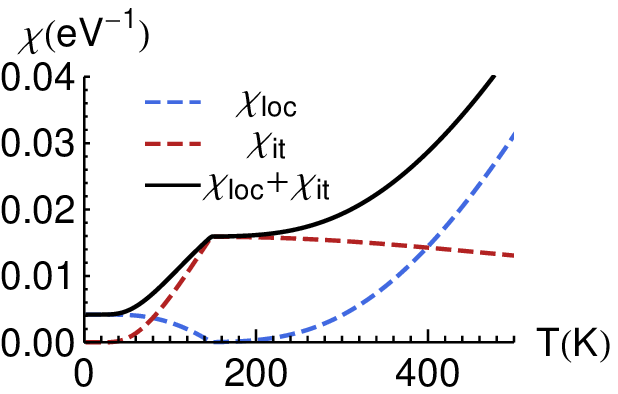}%
\label{fig:ChiComponents}} \quad \subfigure[]{\includegraphics[height=0.11%
\textheight]{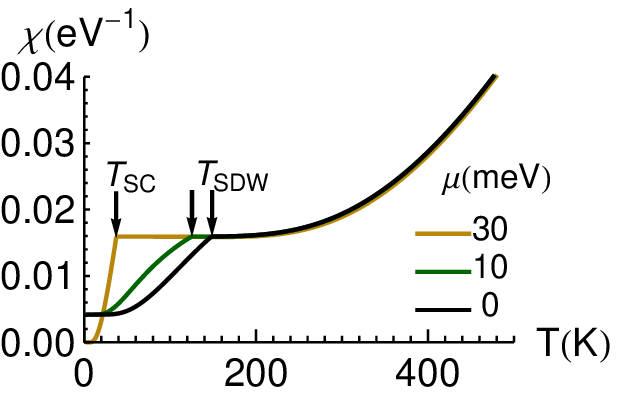}\label{fig:ChiAtDifferentMu}}
\caption{(Color online.) (a) The uniform susceptibility v.s. temperature at $%
\protect \mu=0$. The dashed lines indicate the contributions from either the
local moment (blue) or the itinerant electron (red). The solid line is the
total uniform susceptibility. (b) The total uniform susceptibility curve at
various chemical potentials $\protect \mu$, all shows a rapid drop in the
ordered phase at low temperature.}
\label{fig:Uniform Susceptibility}
\end{figure}

The total uniform susceptibility $\chi =\chi _{\text{loc}}+\chi _{\text{it}}$
at $\mu =0$ is presented in Fig.\thinspace \  \ref{fig:ChiComponents}, in
which the contributions from both the local moment and the itinerant
electron degrees of freedom, i.e., $\chi _{\text{loc}}$ and $\chi _{\text{it}%
}$, are also given, respectively. In Fig.\thinspace \ref%
{fig:ChiAtDifferentMu}, the uniform susceptibility is shown at different $%
\mu $'s where the low-temperature phases are either SDW or SC.

To probe the uniform susceptibility for the local moment, we add a Zeeman
term $-M\sum_{i}h\cdot n_{i}$ to the local moment Hamiltonian, where $h$ is
the uniform magnetic field. Then the local moment action is modified from
Eq.\thinspace \eqref{eq:S_loc} by the replacement $\nu \rightarrow \nu +imh$
with $m=0,\pm 1$ denoting the quantum numbers of the three spin wave modes
respectively. By integrating out the local moment degrees of freedom (i.e. $%
n_{q}$ and $\tilde{n}_{q}$ fields), the free energy for the local moment
reads
\begin{equation}
\begin{split}
F_{\text{loc}}=& \frac{1}{2g}\left( \eta ^{2}-h^{2}\right) n_{0}^{2} \\
+& \sum_{m=0,\pm 1}\sum_{q}\ln ((\nu +imh)^{2}+c^{2}{\bm{q}}^{2}+\eta ^{2}).
\end{split}%
\end{equation}%
Then the uniform susceptibility can be obtained from the second order
derivative $\chi _{\text{loc}}=-\partial ^{2}F_{\text{loc}}/\partial h^{2}$
taken in the $h\rightarrow 0$ limit,
\begin{equation}
\chi _{\text{loc}}=\frac{n_{0}^{2}}{g}+4\sum_{q}\frac{-\nu ^{2}+c^{2}{\bm{q}}%
^{2}+\eta ^{2}}{(\nu ^{2}+c^{2}{\bm{q}}^{2}+\eta ^{2})^{2}}.
\end{equation}%
Carrying out the frequency and momentum summation, we get
\begin{equation}
\chi _{\text{loc}}=\frac{n_{0}^{2}}{g}+\frac{1}{\pi \beta c^{2}}Y(\frac{%
\beta \eta }{2}),
\end{equation}%
where the function $Y(x)=x\coth x-\ln (2\sinh x)$.

In the high temperature limit, according to Eq.\thinspace \  \eqref{eq:eta},
the spin wave mass gap $\eta $ increases linearly with temperature as $\eta
=2k_{B}T$, then the function $Y$ tends to a finite limit $Y(1)=0.458$,
resulting in a linear-$T$ behavior
\begin{equation}
\chi _{\text{loc}}=0.458\frac{k_{B}T}{\pi c^{2}},
\end{equation}%
which will dominate the total uniform susceptibility at high temperature,
consistent with the experiments.\cite{magsus1,magsus2}

On the other hand, the itinerant electron uniform susceptibility can be
evaluated from
\begin{equation}
\chi _{\text{it}}\ =-S^{2}\sum_{k}\text{Tr}\left[ \tau _{i}G(k)\tau
_{i^{\prime }}G(k)\right] ,
\end{equation}%
where $S=1/2$ for the itinerant electrons, and the matrix $\tau _{i}=\sigma
_{0}\otimes \sigma _{i}$ represents the spin operator. The particular form
of the propagator $G(k)$ will depend on the order in the itinerant electron
state. In general the frequency summation involved can be complicated.
However, to the leading order of approximation, we have
\begin{equation}
\chi _{\text{it}}\ =-\frac{1}{2}\sum_{{\bm{k}}}[n_{F}^{\prime }(E_{\bm{k}%
}^{+})+n_{F}^{\prime }(E_{\bm{k}}^{-})]+O(n_{0}^{2},\Delta ^{2}),
\end{equation}%
where $n_{F}^{\prime }$ is the first order derivative of the Fermi
distribution function, and $E_{\bm{k}}^{\pm }$ provides the band structure.
For normal state $E_{\bm{k}}^{\pm }=\pm \epsilon _{\bm{k}}-\mu $, for SDW
state $E_{\bm{k}}^{\pm }=\pm (\epsilon _{\bm{k}%
}^{2}+J_{0}^{2}n_{0}^{2})^{1/2}-\mu $, and for SC state $E_{\bm{k}}^{\pm
}=((\pm \epsilon _{k}-\mu )^{2}+\Delta ^{2})^{1/2}$. The remaining terms are
of the second order of the order parameters. Since the function $%
n_{F}^{\prime }(E_{\bm{k}}^{\pm })$ peaks at $E_{\bm{k}}^{\pm }=0$, so if
the itinerant electron band is gapped from the Fermi surface, the contribution to
the uniform susceptibility will decrease rapidly, which accounts for the
quick drop the total uniform susceptibility in the ordered phase in
Fig.\thinspace \ref{fig:ChiAtDifferentMu}.


\subsection{The renormalization of the itinerant electron band}

\begin{figure}[htb]
\centering
\subfigure[]{\includegraphics[width=0.18\textwidth]{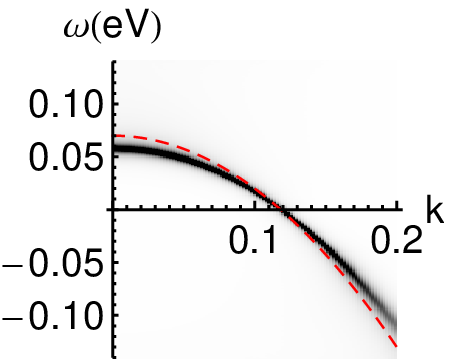}%
\label{fig:ARPES0}} \quad \subfigure[]{\includegraphics[width=0.18%
\textwidth]{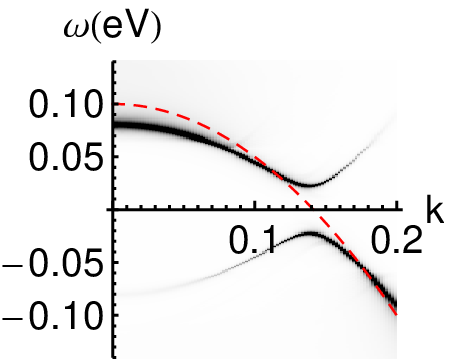}\label{fig:ARPESSDW}}
\caption{(Color online.) The calculated itinerant electron spectrum for the $\Gamma $ pocket
in (a) the normal phase ($\protect \mu =0.03\text{eV}$, $T=70\text{K}$) and
(b) the SDW phase ($\protect \mu =0.00\text{eV}$, $T=70\text{K}$). Darker
shade indicates higher intensity. The red dashed curve marks out the bare
dispersion of itinerant electron.}
\label{fig:ARPES}
\end{figure}

Now we consider the self-energy correction due to the scattering of the
itinerant electrons with local moments, which is given by
\begin{equation}
\Sigma _{e}(k)=-\settowidth{\figwidth}{\includegraphics[scale=0.75]{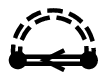}}%
\parbox{\figwidth}{\includegraphics[scale=0.75]{Se.eps}}=J_{0}^{2}\sum_{k^{%
\prime }}D_{ii^{\prime }}(k^{\prime }-k)s_{i}G(k^{\prime })s_{i^{\prime }}.
\end{equation}%
Here the local moment propagator $D$ is taken from Eq.\thinspace \ %
\eqref{eq:D} as the RPA-corrected one, while the bare single-particle
propagator $G$ is given by Eq.\thinspace \eqref{eq:G_0} in the normal state
and Eq.\thinspace \eqref{eq:G_SDW} in the SDW state. The renormalized
single-particle propagator obtained from Dyson's equation
\begin{equation}
\tilde{G}(k)=[G(k)^{-1}-\Sigma _{e}(k)]^{-1}
\end{equation}%
determines the spectral function after a Wick rotation to the real frequency
domain by
\begin{equation}
\tilde{A}(\omega ,{\bm{k}})=-2\mathrm{Im}\tilde{G}(\omega +i0_{+},{\bm{k}}).
\end{equation}%
The result for the hole pocket around the $\Gamma $ point is shown in
Fig.\thinspace \ref{fig:ARPES}. The pocket is slightly more shallow in both
the normal and SDW phases, compared to the bare dispersion as indicated by
the red dashed curve.

\begin{figure}[htb]
\centering
\subfigure[]{\includegraphics[width=0.22\textwidth]{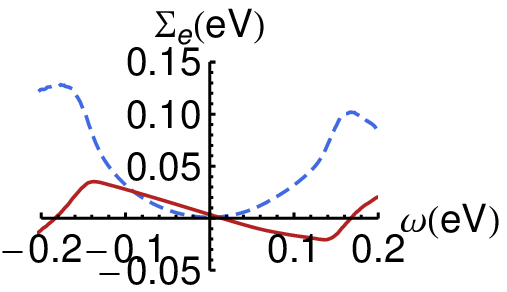}%
\label{fig:SE0}} \quad  \subfigure[]{\includegraphics[width=0.22%
\textwidth]{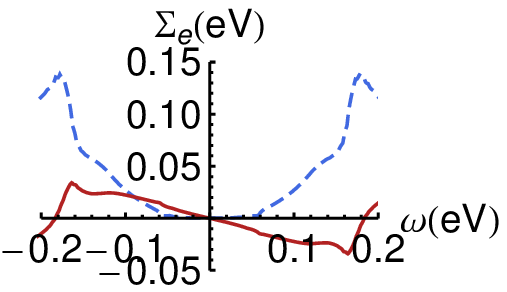}\label{fig:SESDW}}
\caption{(Color online.) The momentum-accumulated self-energy $\Sigma_e$($%
\protect \omega$) in (a) the normal phase ($\protect \mu=0.03\text{eV}$, $T=70%
\text{K}$) and (b) the SDW phase ($\protect \mu=0.00\text{eV}$, $T=70\text{K}$%
). The red solid curve represents the real part $\mathrm{Re}\Sigma_e$, while
the blue dashed curve represents the negative imaginary part $-2\mathrm{Im}%
\Sigma_e$.}
\label{fig:SE}
\end{figure}

This band renormalization effect can be understood by looking at the
frequency dependence of the momentum-accumulated self-energy $%
\Sigma_e(\omega)=\sum_{\bm{k}}\Sigma_e(\omega,{\bm{k}})$, as shown in
Fig.\thinspace \ref{fig:SE}. The negative imaginary part typically has a
valley shape, due to the reduced scattering rate within the local moment gap
$\pm \eta$. It can be well approximated by $-2\mathrm{Im}\Sigma_e(%
\omega+i0_+)\propto \omega^2$ for small frequency $\omega$. According to the
Kramers-Kronig relation, the real part of the self-energy should follow $%
\mathrm{Re}\Sigma_e\propto -\omega$, meaning that the self-energy correction
reduces the electron energy above the Fermi level and increases it below the
Fermi level, thus always squeeze the electron pockets. This partly account
for the reduced pocket depth generally observed in ARPES experiments\cite%
{ARPES1,ARPES2,ARPES3,ARPES4} compared to the LDA calculations\cite%
{SPM1,SPM2,LDA1,LDA4,LDA5}.

\section{Conclusion}

In this paper, we have presented a systematic study of the itinerant
electron and local moment hybrid model\cite{hybrid model,5band} for the
iron-based superconductors. The microscopic origin of both the itinerant
electron and local moment degrees of freedom are all from the $3d$-orbitals
of the iron atoms. As a renormalization flow at low energy, part of the $3d$
electrons is conjectured to form local moments through an orbital-selective
Mott transition. Here as a simplification, a two-pocket band structure is
adopted for the itinerant electrons without considering their Coulomb
interaction. A robust short-ranged AF fluctuations around the momentum ${%
\bm{Q}}_{s}$ is incorporated for the local moment part via a non-linear $%
\sigma $ description tuned in a disordered regime, which can persist up to
high temperature in the normal state. Thus, in this minimal model, an SDW/AF
instability in either subsystem is not intrinsically present when they are
decoupled. A Hund's rule ferromagnetic interaction then couples these two
subsystems together.

What we have established in this work is that such a simple model is
generically infrared-unstable against either magnetic or SC ordering at low
doping, thanks to the \textquotedblleft resonant\textquotedblright \
scattering of the itinerant electrons between the hole-electron pockets by
the local AF fluctuations of the local moments around ${\bm{Q}}_{s}$. In
other words, the itinerant electrons form an SDW/SC order by a strong
coupling to a background AF fluctuations of the preformed local moments with
a momentum match. The phase diagram in Fig. 3 is qualitatively in agreement
with the experimental ones, in which the Cooper pairing is not glued by the
Fermi-surface-nesting driven collective fluctuations of the itinerant
electrons which would otherwise result in a much weaker pairing strength in
a much narrower doping regime, close to the SDW phase, than what has been
shown in Fig. 3. The effective glue provided by the magnetic fluctuations of
the local moment automatically favors the $s^{\pm }$-wave paring symmetry
here. The presence of the local moments further explains the
high-temperature linear-$T$ dependence of the uniform magnetic
susceptibility (Fig. 6) in the normal state. In particular, the strong
scattering between the itinerant and localized electrons is represented by
the dynamic spin susceptibility shown in Fig. 5, which illustrates how the
Goldstone mode in the SDW state becomes a \textquotedblleft
resonant-like\textquotedblright \ mode in the SC state as well as its
evolution in the normal state. The strong signature of the
itinerant/localized electron coexistent picture seen in Fig. 5, including
both low and high energy parts, can serve a very useful qualitative
prediction for the neutron-scattering measurement even if the comparison may
not yet be quantitatively due to the highly simplified nature of the model.

Therefore, the minimal model studied in this paper may be generally used to
describe the low-energy physics in a multiband electron system in which the
electrons in some more localized orbitals may first form short-ranged
(fluctuating) SDW order at a higher characteristic temperature (called the
hidden local SDW order in Ref. \cite{weng}). Then at lower temperatures, the
electrons in more itinerant orbitals can be naturally driven into a true SDW
order or SC state via the Hund's coupling to such a preformed local SDW
background. In contrast to the scenario\cite{Si1} that an electron may carry
both a coherent itinerant and an incoherent local moment signatures, in
analog to a single band case at an intermediate coupling, the multiband case
provides with us an alternative, but simpler possibility, i.e., via the
orbital-selective Mott transition, itinerant and localized electrons may be
explicitly separated as independent degrees of freedom.

\section*{Acknowledgment}

The authors would like to thank X. H. Chen, H. Zhai, T. Li, H. Yao, and P.
Ye for helpful discussions. This work is supported by NSFC grant Nos.
10704008, 10834003 and 10874017 as well as the grants of National Program
for Basic Research of MOST Nos 2011CB921803, 2009CB929402 and 2010CB923003.

\end{document}